\documentclass[11pt]{article}
\usepackage{psfig}
\begin{document}

\title{\LARGE\bf The Power Spectra of Two Classes of
Long-duration 
Gamma-ray Bursts } 

\author{SHEN Rong-feng \ \ \  SONG Li-ming  \\
\it Particle Astrophysics Center, Institute of High Energy
Physics,\\ 
\it Chinese Academy of Sciences, Beijing, 100039}  
\date{} 
\maketitle

\begin{abstract}  

We have studied the averaged power density spectra (PDSs) of two
classes 
of long-duration gamma-ray bursts in the recent classification by
Balastegui 
et al.(2001) based on neural network analysis. Both PDSs follow a
power law 
over a wide frequency range with approximately the same slope,
which 
indicates that a process with a self-similar temporal property
may underlie 
the emission mechanisms of both. The two classes of bursts are
divided into 
groups according to their brightness and spectral hardness
respectively and 
each group's PDS was calculated;  For both classes, the PDS is
found to 
flatten both with increasing burst brightness and with increasing
hardness.

\end{abstract}


\section*{1. INTRODUCTION}

Gamma-Ray Bursts (GRB) are transient and violent phenomena in
which an energy of 
$\sim 10^{51}-10^{53}$ ergs is released in a few seconds, in the
form of 
$\gamma$ rays. 
They were first discovered 
by the USA Vela satellite in 1967, 
but 
we
still know very little about their nature.
The BATSE (Burst and Transient Source Experiments) on board the
Compton Gamma-
Ray Observatory (CGRO) of USA 
launched in 1991, 
has since observed over 2000 GRBs.
These bursts show
a great 
diversity 
as regards 
duration, peak flux and 
form of the energy spectrum, as well as 
time profile. This indicates that the 
process of GRB generation 
may be very complex. Meanwhile, 
the time profiles
are richly
packed 
with
information 
on the GRB mechanism, 
and the study of their temporal properties 
is a valuable path to their understanding.

Beloborodov et al.(1998, 2000) calculated the average Fourier
power density 
spectrum (PDS) of some bright and long 
gamma-ray 
bursts. They found the average PDS follow a power law over almost
2 decades of 
Fourier frequencies with two deviations (breaks), 
one 
at the low frequency end and 
one at the high frequency end. The low frequency break is caused
by the finite durations of the bursts, i.e., the ``window
effect''; the high 
frequency break at 1$\sim$2 Hz can be related to a typical time
scale of 
0.5$\sim$1 s. Variabilities taking place on time scales shorter
than that 
either 
do not exist,
or 
are suppressed by some mechanism. Power-law fitting to the
average PDS
gives an 
index of -5/3, the same slope of the Kolmogorov spectrum
describing velocity 
fluctuations in a turbulent medium. These results 
suggest the presence of a physical process which 
is 
self-similar 
over a wide range of
time scales, possibly 
turbulence in magnetic field 
(Stern, 1999).

A 
number of models have been proposed to explain the ``central
engine'' of GRBs, 
merger of 
two neutron stars,
of a neutron star 
and a black hole,
collapse of a massive star, rapidly spinning and strongly
magnetized compact 
objects, phase transition of compact objects, 
accretion onto massive black holes, and so on (Cheng \& Lu,
2001). Do all GRBs 
have the same origin and mechanism\ ? Or can
they be divided into different classes 
according to origin\ ? This is an attractive problem.

The 
best known classification is the 
classification 
into 
a short-duration class (T$_{90} <$ 2 s)
and 
a long-duration class (T$_{90} >$ 2 s), 
$T_{90}$ 
being the interval between 
the times 
of 
5\% and 95\% 
cumulative 
photon
counts. The short bursts have harder energy spectra, and the long
bursts, softer 
ones (as represented by the hardness ratio H$_{32}$) (Kouveliotou
et al., 
1993). Some people 
tried a multivariate classification 
with the duration, hardness, brightness, etc., as variates
(e.g., Mukherjee et al., 1998, Balastegui et al., 2001).
Balastegui et al.\ 
(2001) 
applied 
neural network analysis 
and 
divided 1599 bursts 
into 3 classes: Class I (531 bursts) corresponds to the ``short''
class of the
above classic classification, and  Class II (341) and class III
(727) 
together correspond to the ``long'' class. 
See Fig.\,1.
The authors suggest that the two new classes of long bursts may
correspond to 
different progenitors: Class II 
to mergers of 
two neutron stars,
or a
neutron star 
and a black hole; Class III 
to collapse of massive stars.     

\begin{figure}
\centerline{\psfig{file=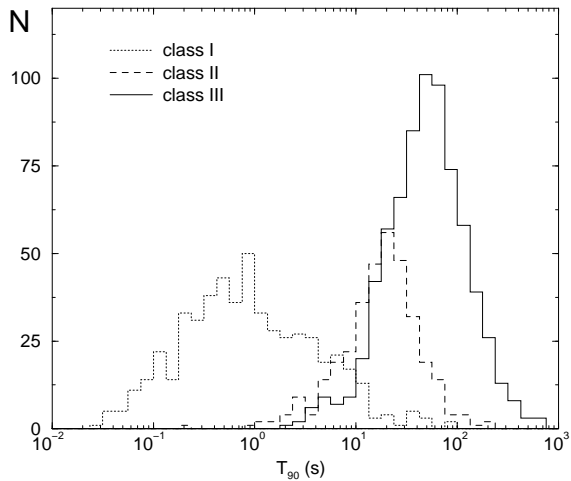,width=7.5cm}}
\caption{\small Duration distributions of the three-class
classification 
from neural network analysis (from Balastegui et al.,2001)}
\end{figure}

In this paper, we try to analyze the 
results of 
Balastegui's 
classification 
as regards 
their average PDS.  
Different from the bulk quantities of GRBs, the PDS can depict
temporal 
properties of the whole light curve, and may reflect information
that is 
directly related to the physical process of 
their 
generation. Therefore, 
a PDS investigation can 
serve
as an independent test of the reclassification. 
Now, 
most of Class I have 
too short duration
for 
the Fourier PDS calculation; 
moreover, 
many
lines
of evidence 
point
to 
the T$_{90} <$2\,s 
bursts  
belonging
to a different class 
of objects 
(Qin et al., 2000; Hakkila et al., 2000). So only bursts of Class
II and Class 
III are included in our investigation.

\section*{2. CALCULATION OF AVERAGE PDS}

For every individual burst, we use concatenated 64ms data summed
over energy 
channel 2 (50$\sim$100 KeV) and channel 3 (100$\sim$300 KeV),
recorded by the 
LAD detectors of BATSE. The average PDS is calculated 
according to
the following procedure:

(1) Select 
bright and long bursts. For dim bursts, 
the low signal to noise ratios will reduce the reliability of
calculated 
results; and for bursts of too short durations, the ``window
effect'' will 
severely distort the shape of 
the PDS. From Balastegui et al.\ (2001)'s sample, we 
therefore
selected bursts 
with T$_{90}>$ 10 s
and with 
$P_{1024}>0.4$ photons cm$^{-2}$ s$^{-1}$ ($P_{1024}$ is the peak
flux recorded 
with 1024 ms time resolution, 
and 
is taken as a 
measure of the brightness). 

(2) Calculate 
the PDS of individual burst. 
First,
the background is subtracted from the light curve. We 
then 
calculate the Fourier transform of each light curve, using the
standard Fast 
Fourier Transform method. 
The power density at frequency $f$
is 
the squared amplitude of 
the Fourier transform at $f$. The power density of 
Poisson 
noise 
is equal 
to the total photon 
count of the light curve including the background. We calculate
the individual 
Poisson noise power density and subtract it from the burst PDS.

(3) Averaging of PDS. Normalization of the individual PDS is
needed before the 
averaging. We use the peak-normalization procedure: 
we determine the peak count rate, $C_{peak}$ $[$photons per 64
ms$]$, with 
the background subtracted, and normalize the individual PDS by
$C_{peak}^2$. The 
average PDS is then calculated by summing up the 
normalized PDSs and dividing 
by the number of bursts.

\section*{3. RESULTS AND ANALYSIS}

The resulting average PDSs of the two classes of bursts are
plotted in Fig. 2. 
Both show a power-law 
form, $P_f \propto f^{\alpha}$, 
over a frequency range of 
more than one decade. There are deviations (breaks) from the
power-law at the 
low frequency end and the high frequency end. 
The high frequency end 
break for  
the Class II bursts 
is not well determined 
because of 
the small 
sample size. 
The locations of 
the low frequency break for the two classes are different,
because they have 
different average 
burst durations. 
In the case of peak-normalization, 
a burst with a longer duration
have a greater 
fluence, and consequently 
gives a larger Fourier power. Since 
the Class III 
objects have,
on average, 
longer durations, 
their PDS curve 
consequently 
lies above 
that of the Class II objects, 
as is shown in Fig.\,2.

\begin{figure}
\centerline{\psfig{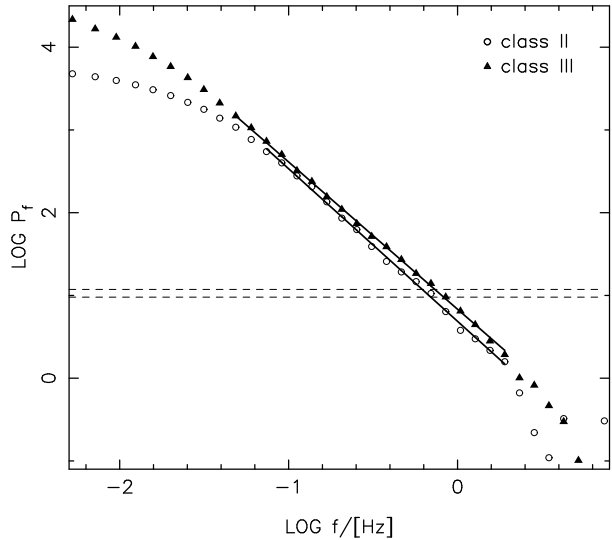}}
\caption{\small The average PDSs for class II and class III 
from neural network analysis. The horizontal dashed lines 
show the averaged noise power density levels of two classes. 
The solid lines are the pwer-law fits.}
\end{figure}

We fitted the average PDSs of 
the
two classes 
with the power law $P_f \propto f^{\alpha}$ 
using the $\chi^2$ fitting procedure. The fitting results are
listed in Table 1; 
also listed are the average values of 
the duration T$_{90}$, 
peak flux $P_{1024}$ and 
hardness ratio H$_{32}$ (defined as the 
fluence ratio of 
of channel 3 to 
channel 2). The fact that 
the 
power-law index $\alpha$ of the average PDS decreases with the
average burst 
brightness was found by Beloborodov et al.(2000), and 
this finding 
is confirmed by our results (see below). 
Because our sample includes 
a greater number of
dimmer bursts,
the power-law index of 
our average PDS 
is less  than -5/3.

\begin{table}[h] \small 
\begin{center}
\caption{\small The average PDSs for class II and class III from 
neural network analysis.}
\vspace{.3cm}
\begin{tabular}{cccccc} \hline
Class & Number $^*$ & $\overline{\rm T}_{90}$\,/s &
$\overline{\rm P}_{1024}$\,/ph\,cm$^{-2}$s$^{-1}$ &
$\overline{\rm H}_{32}$ & $\alpha$ \\ \hline
II  & 206       & 28.9  & 0.92 & 3.03 & -1.84   \\
III  & 650      & 74.3  & 4.28 & 3.15 & -1.78   \\ \hline
\end{tabular} \\
\hspace{-5cm} $*$ T$_{90}>$10\,s,
P$_{1024}>$0.4\,ph\,cm$^{-2}$s$^{-1}$
\end{center} 
\end{table}

We divided 
the bursts of the two classes into groups 
of 
approximately 
the same size,
according to their peak fluxes ($P_{1024}$) and 
to their hardness ratios (H$_{32}$), respectively.
For each group
we calculated its average PDS,
which in every case showed a power law form. 
Their power-law indices $\alpha$,  
are plotted against 
$P_{1024}$ and H$_{32}$, 
in 
the two panels of Fig. 3. 
We see that, for both Class II and Class III objects,
the average PDS flattens 
both with increasing peak 
brightness 
and 
with increasing hardness ratio.

\begin{figure}
\centerline{\psfig{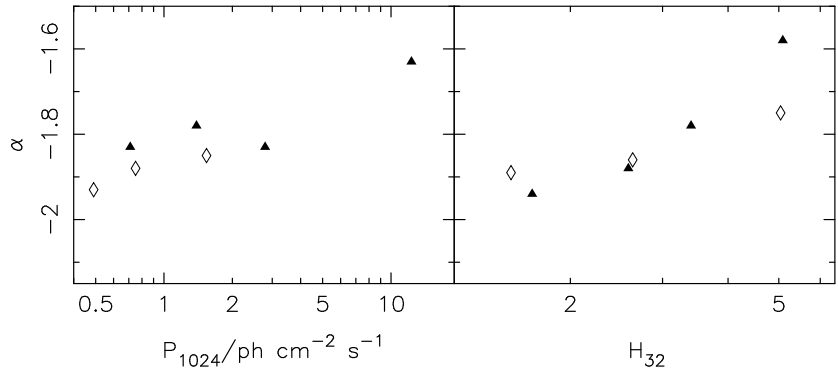}}
\caption{\small The average PDSs for groups of class II and class
III 
from neural network analysis: $\alpha$ vs. P$_{1024}$ and
$\alpha$ 
vs. H$_{32}$. {\Large $\diamond$} --- class II, $\triangle$ ---
class III.}
\end{figure}

Flattening of the average PDS 
implies 
an increase of the components
of 
short time-scale variabilities. 
Now, 
GRBs are usually supposed to be at cosmological distances; dimmer
bursts are 
more distant. Given that different 
(brightness) groups of bursts are at different cosmological
distances, the 
question arises, 
could  the decrease in 
$\alpha$ 
be caused by cosmological time dilation ?  
But time dilation stretches the 
entire light curve: 
for any given spectral structure  
its Fourier frequency is lowered by a factor independent of 
that frequency. Therefore, if the PDS is a power-law, 
time dilation will {\it not\/} change 
its slope. 
We therefore conclude
that the increasing short time-scale variabilities, 
as indicated by the flattening average PDS, reflects an evolution
of temporal 
properties of some intrinsic dynamic process of the GRBs.

\section*{4. DISCUSSION}

As 
shown by our calculation, the two classes of long GRBs, 
in the new 
classification 
of Balastegui et al.\ (2001), do not show 
any notable difference as 
regards the 
shape and 
structure of their average PDSs, except for some trivial
differences  between 
their fitted power-law indices. The power-law form of 
the 
average PDS indicates that a 
physical process which 
has
self-similarity 
over a wide time-scale range may 
be at work 
in both 
two classes. 
This process 
was supposed by some 
to be 
turbulence in magnetic field (Stern, 1999). Many 
different models 
of the progenitors of GRBs can give rise to the 
required 
``fire ball''; Balastegui et al.(2001) even claimed that the
three classes of 
GRBs 
in their classification 
correspond to three kinds of progenitors. But our results about 
the
average PDSs of the two long GRB classes calculated directly from
the GRB light 
curves indicate that the light curves are not sensitive to the
progenitors, even 
if the two classes originate 
in
different progenitors. 
It 
is probable that both 
classes undergo a 
phase 
of relativistic expanding ``fire ball'', in which energy is
dissipated via 
magnetic reconnection explosion caused by turbulence (Stern,
1999). After the 
``fire ball'' 
phase, information about the progenitor 
is lost. On the other hand, if GRB light curves are supposed to
contain 
information about the progenitor, 
since some theoretical models of GRBs do not essentially need the
``fire ball'' 
phase, 
then our results reduces the probability that these long bursts
have different 
origins.  

Beloborodov et al.(2000) and Pozanenko et al.(2000) found that 
the
average PDS  
flattens with increasing 
burst 
brightness. Under the assumption of cosmological origin and 
a 
``standard candle'', i.e., brighter bursts have smaller
redshifts, 
and also assuming all bursts have the same mechanism, we 
saw 
that
the redshift will not alter the shape of 
the
average PDS (section 3). We conclude that the 
trend of 
the average PDS flattening with 
increasing 
brightness indicates that there exists an evolution of sources,
at least an 
evolution 
in the temporal properties of 
the 
generation processes of GRBs. 

The other trend of 
the
average PDS flattening with 
increasing hardness ratios may be associated with another 
observational fact --- the pulses in 
the 
GRB's light curve 
are 
narrower in higher energy channel (the pulse is 
a 
fundamental morphological 
entity 
of 
the
GRB light curve) (Norris et al., 1996). Furthermore, Beloborodov
et al.(2000) 
showed that 
the
average PDS 
gets 
flatter from energy channel 1 to channel 4. The trend we obtained
that 
the
average PDS flattens with hardness ratios is consistent with
their result.\\ 

ACKNOWLEDGEMENT~~ The authors are grateful to Dr. Andreu
Balastegui who 
have provided detailed results of GRB classification using neural
network 
analysis.

\end{document}